# Magnetotransport and Domain Wall in Nanoconstriction of Ferromagnetic Semiconductor (Ga,Mn)As

T. Figielski[1], T. Wosinski[1], O. Pelya[1], J. Sadowski[1,2], A. Morawski[1], A. Makosa[1], W. Dobrowolski[1], R. Szymczak[1], and J. Wrobel[1]

[1]Institute of Physics, Polish Academy of Sciences, 02-668 Warszawa, Poland
[2]Max-lab, Lund University, 22100 Lund, Sweden

We studied magnetoresistance (MR) of nanoconstrictions created in (Ga,Mn)As epilayers by $O^+$ ion implantation. Original layers exhibit a negative MR that is plausibly caused by weak localization (WL) effects at the lowest temperatures. In constricted samples, additionally, jumps of an enhanced conductance appear on the background of the negative MR, whose positions reflect the hysteresis of magnetization. We argue that they are manifestation of a suppression of WL due to the nucleation of a domain wall in the constriction.

The interplay between electron transport through domain walls (DWs) and magnetic properties of ferromagnetic nanowires and point contacts became a subject of a great current interest. This interest is stimulated by at least two reasons. On one hand, a giant magnetoresistance (MR), up to several hundred percent at room temperature, has been observed for Ni, Co and Fe point contacts [1,2]. This has opened potential for their application in highly compacted magnetic-field sensors and read heads. On the other hand, it has been discovered that the dynamics of single domain walls in ferromagnetic nanowires can be effectively studied utilizing the MR effect [3-5]. In turn, recent advance in the growth of ferromagnetic semiconductors based on III-V compounds brought about a possibility for integrating electronic and magnetoelectronic devices. Films of $Ga_{1-x}Mn_xAs$ containing a few percent of Mn atoms can be grown by a low-temperature molecular-beam epitaxy (LT-MBE) [6]. They are *p*-type semiconductors, with a hole concentration being a fraction of the Mn concentration. Below a transition temperature, $T_C$, they become ferromagnets due to a hole-mediated ordering of the Mn spins [7].

We have investigated nanoconstrictions fabricated in a ferromagnetic $Ga_{0.99}Mn_{0.01}As$ film grown by the LT-MBE on a semi-insulating GaAs (001) substrate. The film was 50 nm thick and was covered with a 10 nm-thick GaAs cap layer (see Ref. [8] for more details). Magnetic properties were measured using a superconducting quantum interference device (SQUID) magnetometer with a magnetic field applied parallel to the film plane. Magnetization of the film as a function of temperature, after subtraction of diamagnetic contribution of the GaAs substrate, shows the onset of ferromagnetic ordering at a temperature of 50 K; Fig. 1. Its shape differs from the universal curve for ferromagnets, what could evidence for a magnetic non-uniformity of the film. The film exhibits metallic-type hole conductivity. Individual samples of an outline 2.5×1 mm$^2$ were defined in the film and their distant terminals were supplied with Ohmic contacts. Sample conductance was measured using pseudo-four-probe method and lock-in technique with a sensing voltage of a few mV at 770 Hz, and at a selected dc bias.

The ferromagnetic samples investigated by us exhibit a negative magnetoresistance both for the perpendicular (Fig. 2) and in-plane (perpendicular to the current) orientation of a magnetic field with respect to the film plane. Additionally, for the perpendicular orientation, a small rise of conductance appears at the lowest temperatures in a narrow range of the fields around $H = 0$. Such a behavior of MR is typical for (Ga,Mn)As films. An isotropic negative MR has been commonly attributed to the reduction of spin-disorder carrier scattering due to aligning of localized Mn spins by an external field. However, this explanation holds well only for temperatures above $T_C$ [6,9] and fails at lower temperatures. Another contribution to the observed MR, which possibly dominates at the lowest temperatures, is the suppression of carrier localization by an external magnetic field [10-12].

We first fabricated narrow constrictions in the (Ga,Mn)As film by an electron-beam lithography and chemical etching [13]. However, the samples with constrictions of submicron width, obtained in this way, were non-conducting at liquid helium temperatures, due presumably to a trapping of charge carriers in surface states appearing on an extra area denuded by the etching. Photoconductance in those samples could be excited by illumination generating electron-hole pairs, but its magnitude was strongly suppressed by an applied magnetic field, indicating at some field-dependent recombination process.

Since the chemical etching turned out to be useless for the tailoring of conducting nanoconstrictions in our (Ga,Mn)As films, we next applied for this aim an ion implantation. We have found that implantation of oxygen ions, $O^+$, with energy of 25 keV and a dose as low as $5 \times 10^{13}$ ions/cm$^2$ destroys both the conductivity and ferromagnetism in the film [13]. Mechanism of this destruction is not fully understood. We implanted oxygen ions into the film through a mask that consisted of a 500 nm-thick PMMA resist deposited on the top of the film, and contained windows patterned by help of electron-beam lithography. In this way we fabricated constrictions with lithographic widths of 0.5 and 0.75 μm.

Conductance of the constricted samples has been reduced by two orders of magnitude as compared with that of reference (non-constricted) ones. This conductance rises slightly with temperature in the investigated range of 1.5 – 13 K, and depends on the bias voltage, displaying a distinct minimum at zero bias. The samples show a negative MR typical for the (Ga,Mn)As films; Fig. 2. What is the most interesting, and makes the essence of this Letter, there are abrupt jumps of an enhanced conductance appearing in the constricted samples for the both field orientations; Figs 3 and 4. Positions of these jumps reflect the hysteresis loop of magnetization. Surprisingly, for the in-plane field orientation the two distinct jumps, corresponding to the field swept in opposite directions, merge into a single jump. With increasing temperature the amplitude and extension of the jumps are suppressed and the jumps practically disappear at 13 K. Small irregular features accompany each jump on its both sides and on its top, which are reproducible in subsequent field sweeps but not in subsequent cooling cycles.

In the following we assume that the MR at the lowest temperatures is due to weak localization in highly disordered films of (Ga,Mn)As. Basically, weak localization (WL) arises due to the constructive interference between pairs of wave packets corresponding to an electron traveling diffusively along a closed trajectory in opposite directions. That interference enhances electron backscattering and thus decreases conductivity. Magnitude of WL correction to the conductivity is limited by the time of phase coherence of the two interfering waves, which is determined by the processes of inelastic and spin-flip scattering. In non-magnetic materials, a strong spin-orbit scattering can alter the constructive interference into the destructive one, leading to the so-called antilocalization. Instead, in ferromagnets processes leading to the antilocalization are quenched by the exchange interaction [14]. It is crucial here that a magnetic flux bounded by a closed path introduces a phase difference between the time-reversed paths, which suppresses WL giving rise to an apparent negative MR.

The film investigated is characteristic of a high disorder and falls into the category of dirty metals. The product $k_F l_e$ ($l_e$ is the hole mean elastic free path, and $k_F$ is the Fermi wave number) is of the order of unity at liquid helium temperatures. In that case the semiclassical picture of charge transport fails. However, as shown by Minkov *et al*. [15], the conductivity may then be still treated as quasi-diffusive with corrections due mainly to the quantum interference. Weak localization can coexist with ferromagnetism in our samples. The maximum value of an internal magnetic induction, $B_{int} \approx \mu_0 M_s = 0.012$ T, estimated from the magnitude of saturation magnetization $M_s$, is small enough to be generally neglected in a WL correction to conductivity, with an exception that will be discussed in the following.

To describe the low-temperature MR of our film, we use the WL theory developed by Dugaev *et al*. [14] for 2D ferromagnetic systems in a perpendicular magnetic field. A change of WL correction to conductivity in dependence on a magnetic induction, $B$, is given by:

$$\Delta\sigma(B) = -\frac{e^2}{4\pi^2\hbar} \sum_{\sigma=\uparrow,\downarrow}\left[\Psi\left(\frac{1}{2} + \frac{B_e^\sigma + B_{so}^\sigma}{B}\right) - \Psi\left(\frac{1}{2} + \frac{B_i^\sigma + 2B_{so}^\sigma}{B}\right)\right],$$

where the summation is over the two spin orientations, $\Psi(x)$ is the digamma function, and the parameters $B_n$ (in the following we assume $B_n^\uparrow = B_n^\downarrow = B_n$) are defined by the scattering times $\tau_n$ as: $B_n = \hbar/4eD\tau_n$, where $D$ is the diffusion coefficient. The indices $n$ stand for the following scattering processes: $e$ – elastic, $i$ - inelastic, $so$ – spin-orbit (without the spin flip). When the elastic time is much shorter than other scattering times involved, then $\Delta\sigma(B)$ is practically independent of $B_e$. Thus we have only two fitting parameters in the theory: $B_i$ and $B_{so}$, and possibly a geometrical factor linking the sample conductance with their conductivity. In practice, the main parameter that actually determines the overall shape of $\Delta\sigma(B)$ curve is $B_{so}$. With the fitting parameters: $B_{so} = 0.185$ T, $B_i = 0.02$ T, or equivalently: $(D\tau_{so})^{1/2} = 30$ nm, $(D\tau_i)^{1/2} = 90$ nm, we obtain a fine agreement between the 2D WL theory and the experiment, as seen in Fig. 2. These obtained values show, however, that our samples fall rather into a regime intermediate between the 2D and 3D localization.

The small rise of conductance appearing around $H = 0$ (see Fig. 4) requires a separate explanation. It has been commonly attributed to the reorientation of magnetization from the in-plane direction to perpendicular direction [9,11]. We propose that WL is directly responsible for this effect. The in-plane spontaneous magnetic induction in the film exceeds the perpendicular induction after rotation of magnetization, because of a large difference in the demagnetizing factors for the both orientations. Therefore the negative localization correction to conductivity would be more suppressed by an internal magnetic flux for the in-plane magnetization direction than for the perpendicular direction, what satisfactorily explains the near-zero-field feature. A possible difference in the spin-orbit scattering times for different magnetization directions, pointed out in Ref. [14], seems to be inessential here.

To discuss the origin of the extra conductance jumps observed in the constricted samples, we have to take into account an effect of domain wall on the electron transport. It is essential that a DW tends to localize itself in the constriction, in order to minimize its energy. As pointed out by Bruno [16], a geometrically constrained wall differs from that in a bulk ferromagnet. It is neither Bloch nor Néel wall and its width is essentially determined by the constriction size. Several authors studied theoretically an effect of electron scattering at domain walls on resistivity of ferromagnetic metals [17-19]. Theories predict generally a large *positive* contribution of DW to the resistivity if the wall is thin enough compared to the Fermi wavelength of conducting charges, and if the charge transport through the wall is ballistic.

Instead, we observe in the constricted samples a distinct increase in conductance when a domain wall is nucleated in the constriction. This points out a *negative* contribution of DW to the resistivity. It is intelligible in the framework of localization theory. Tatara and Fukuyama [20] predicted that a DW destroys the electron phase coherence necessary for weak localization in the case of a dirty metal, what leads to a negative contribution of DW to the resistivity. Also, Lyanda-Geller *et al.* [21], Takane and Koyama [22] and Jonkers [23] arrived at similar conclusions. Basically, the predicted effect is due to misalignment of the localized spins inside the wall. Authors of the pioneering work [20] conclude with the following statement: "It will be interesting to observe in magnetic wires this reduction of resistivity by the nucleation of domains in more definite ways". We believe that our experiments just supply with such an observation.

Thus, the appearance of the jumps of an enhanced conductance can be understood as due to erasing of the quantum localization by a DW nucleated in the constriction. Each jump would extend over the field range, in which the DW is pinned at the constriction. Small features accompanying each jump are probably associated with changes in the domain structure of the sample outside the constriction. This our finding contrasts with jumps of a reduced conductance observed recently by Rüster *et al.* [24] for very narrow constrictions in $Ga_{0.976}Mn_{0.024}As$ epilayers. We attribute this divergence to the difference in constriction width of the samples used in the both cases, which had to be followed by a difference in the domain width. Thus, while the charge transport through a DW in our samples is diffusive, it is likely ballistic (or tunnel) for the samples studied in Ref. [24].

Coming to an end, we want to pay attention to a particular mechanism that may contribute to the erasing of WL by a DW. Assume that an itinerant spin follows adiabatically the direction of local magnetic induction $\mathbf{B}_{int}$ inside a DW. Then, the wave function of a spin-carrying particle, which makes a closed loop to return to its starting point, acquires an additional geometrical Berry-phase [21]. For a spin-1/2 particle this geometrical phase equals half of the solid angle $\Omega$ subtended by the vector $\mathbf{B}_{int}$ experienced by the traveling particle, with a sign depending on the spin orientation and the direction of circulation. In a geometrically constrained DW, whose width is comparable to the constriction width, the direction of rotating magnetization may vary along two lateral dimensions instead of one as in the case of Bloch or Neel wall [16]. For instance, we can imagine ourselves that a DW has a mixed Bloch-Néel character with a relative contribution of each component changing across the constriction width. Then, for closed trajectories running inside the DW the solid angle $\Omega$ subtended by $\mathbf{B}_{int}$ will be different from zero, giving rise to a non-zero geometrical-phase difference between the time-reversed paths. That difference tends to destroy the constructive interference and to erase the localization effects.

In conclusion, we argue that the low-temperature magnetoresistance of ferromagnetic (Ga,Mn)As epilayers results predominately from the suppression of weak localization in a magnetic field. Our experiments performed with nanoconstricted samples demonstrate that nucleation of a domain wall in the constriction erases the effects of weak localization, leading to a negative contribution of the wall to resistivity.

We are indebted to M. Baran[1] and J. Jagielski (Institute of Electronic Materials Technology, Warsaw) for their assistance. This work has been partly supported by the Committee for Scientific Research of Poland under Grant No. 1 P03B 052 26.

[1] N. García, M. Muñoz, and Y.-W. Zhao, Phys. Rev. Lett. **82**, 2923 (1999); Appl. Phys. Lett. **76**, 2586 (2000).

[2] N. García, G.G. Qiang, and I.G. Saveliev, Appl. Phys. Lett. **80**, 1785 (2002).

[3] R. Danneau *et al.*, Phys. Rev. Lett. **88**, 157201 (2002).

[4] R. Hanada *et al.*, J. Phys.-Cond. Mat. **14**, 6491 (2002).

[5] H.X. Tang *et al.*, Phys. Rev. Lett. **90**, 107201 (2003).

[6] F. Matsukura *et al.*, Phys. Rev. B **57**, R2037 (1998).

[7] T. Dietl *et al.*, Science **287**, 1019 (2000).

[8] J. Sadowski *et al.*, J. Vac. Sci. Technol. B **18**, 1697 (2000).

[9] A. Oiwa *et al.*, Sol. State Commun. **103**, 209 (1997).

[10] T. Figielski *et al.*, Acta Phys. Pol. A **103**, 525 (2003).

[11] K.W. Edmonds *et al.*, J Appl Phys. **93**, 6787 (2003).

[12] T. Dietl *et al.* in Recent Trends of Physical Phenomena in High Magnetic Fields, ed. D. Vagner *et al.*, Kluver Academic Publishers, 2003, p.197.

[13] T. Figielski *et al.*, Phys. Stat. Sol. (a) **195**, 228 (2003).

[14] V.K. Dugaev, P. Bruno, and J. Barnaś, Phys. Rev. B, **64**, 144423 (2001).

[15] G.M. Minkov *et al.*, Phys. Rev. B, **65**, 235322 (2002).

[16] P. Bruno, Phys. Rev. Lett. **83**, 2425 (1999).

[17] G.G. Cabrera and L.M. Falicov, Phys. Stat. Sol. (b) **61**, 539 (1974).

[18] L.R. Tagirov, B.P. Vodopyanov, and K.B. Efetov, Phys. Rev. B **63**, 104428 (2001).

[19] M. Viret *et al.*, Phys. Rev. B **53**, 8464 (1996).

[20] G. Tatara and H. Fukuyama, Phys. Rev. Lett. **78**, 3773 (1997).

[21] Y. Lyanda-Geller, I.L. Aleiner, and P.M. Goldbart, Phys. Rev. Lett. **81**, 3215 (1998).

[22] Y. Takane and Y. Koyama, J. Phys. Soc. Jap. **69**, 328 (2000).

[23] P.A.E. Jonkers, J. Magn. & Magn. Mat. **247**, 178 (2001).

[24] C. Rüster *et al.*, Phys. Rev. Lett. **91**, 216602 (2003).

# Figure captions

FIG. 1. In-plane magnetization (in units of Bohr magnetons per Mn atom) vs. temperature for $Ga_{0.99}Mn_{0.01}As$ epilayer after subtraction of diamagnetic contribution from the substrate. Inset: magnetization hysteresis loop at 5 K.

FIG. 2. Conductance (in units of $e^2/h$) vs. magnetic field perpendicular to the film plane for $Ga_{0.99}Mn_{0.01}As$ epilayer: (a) reference sample at 1.5 K: full line – experimental curve, open circles – fitted values calculated on the basis of weak localization theory [14] – left scale; (b), (c) and (d) constricted sample at temperatures 13 K, 4.5 K and 1.5 K – right scale. Inset: constricted sample design.

FIG. 3. Change in conductance (in units of $e^2/h$) for the constricted sample, same as in Fig. 2(b-d), vs. magnetic field swept in opposite directions (differentiated by circles or triangles), applied either perpendicular (a) or parallel (b) to the film plane.

FIG. 4. Same as in Fig. 3(a) but for a broader range of sweeping field, showing characteristic rise of conductance near zero field in addition to conductance jumps.

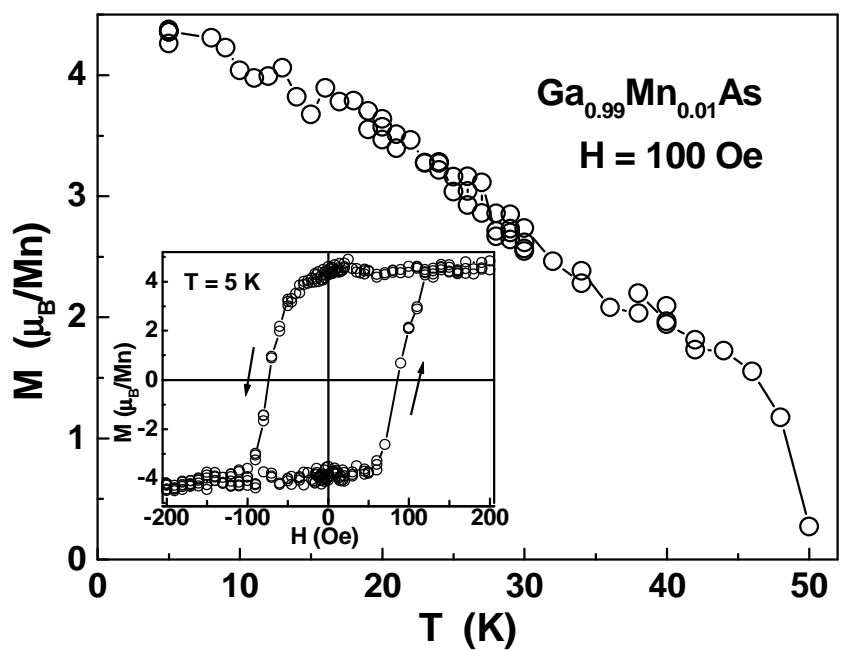

Fig. 1.

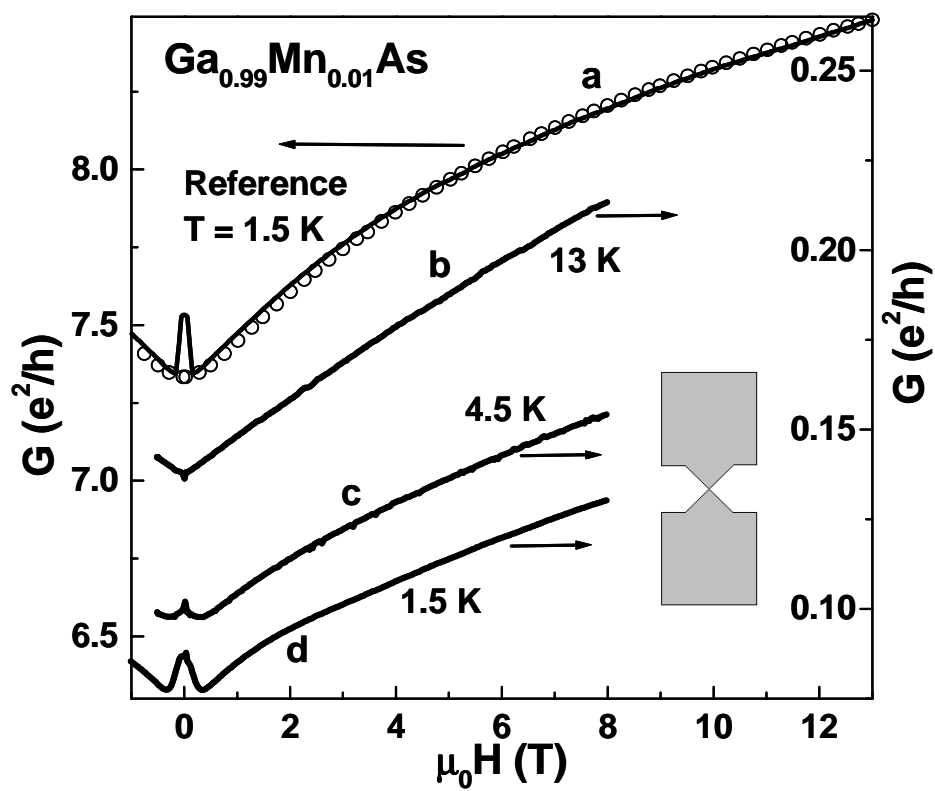

Fig. 2.

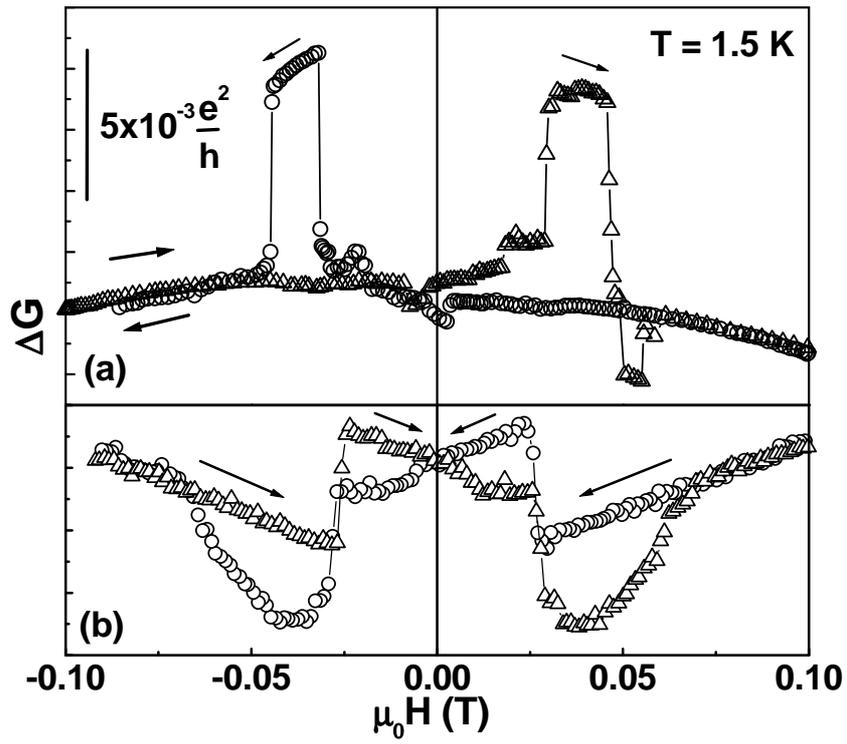

Fig. 3.

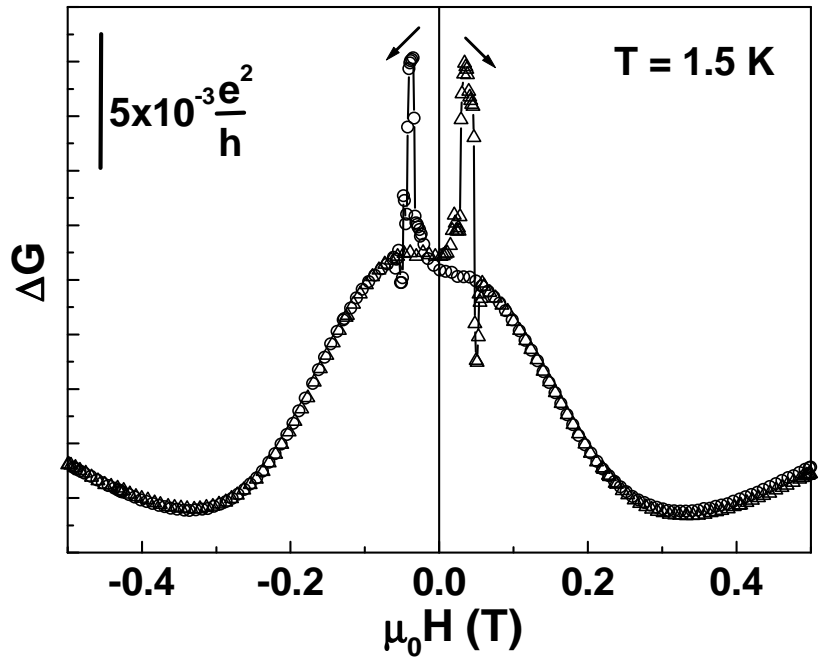

Fig. 4.